\documentclass[twocolumn,aps,prb,showpacs,raggedbottom,nobalancelastpage,amssymb,superscriptaddress]{revtex4-1}

\usepackage{graphicx}
\usepackage{amsmath}
\usepackage{amssymb}
\usepackage{bm}
\usepackage[usenames]{color}
\usepackage{amsfonts}
\usepackage{appendix}
\usepackage{dsfont}
\usepackage{color}
\usepackage{braket}

\begin{document}

\title{Non-quantized square-root topological insulators:\\a realization in photonic Aharonov-Bohm cages}
\author{Mark Kremer}
\thanks{These two authors contributed equally.}
\affiliation{Institut f\"ur Physik, Universit\"at Rostock, Albert-Einstein-Strasse 23a, 18059 Rostock, Germany.}
\author{Ioannis Petrides}
\thanks{These two authors contributed equally.}
\affiliation{Institut f\"ur Theoretische Physik, ETH Z\"{u}rich, Wolfgang-Pauli-Stra\ss e 27, 8093 Z\"urich, Switzerland.}
\author{Eric Meyer}
\affiliation{Institut f\"ur Physik, Universit\"at Rostock, Albert-Einstein-Strasse 23a, 18059 Rostock, Germany.}
\author{Matthias Heinrich}
\affiliation{Institut f\"ur Physik, Universit\"at Rostock, Albert-Einstein-Strasse 23a, 18059 Rostock, Germany.}
\author{Oded Zilberberg}
\affiliation{Institut f\"ur Theoretische Physik, ETH Z\"{u}rich, Wolfgang-Pauli-Stra\ss e 27, 8093 Z\"urich, Switzerland.}
\author{Alexander Szameit}
\affiliation{Institut f\"ur Physik, Universit\"at Rostock, Albert-Einstein-Strasse 23a, 18059 Rostock, Germany.}
\begin{abstract}
Topological Insulators are a novel state of matter where spectral bands are characterized by quantized topological invariants. This unique quantized non-local property commonly manifests through exotic bulk phenomena and corresponding robust boundary effects. In our work, we report a new type of topological insulator exhibiting spectral bands with non-quantized topological properties, but with a quantization that arises in a corresponding system where the square of the Hamiltonian is taken. We provide a thorough theoretical analysis as well as an experimental demonstration based on photonic Aharonov-Bohm cages to highlight the bulk and boundary properties of this neophyte state of matter.
\end{abstract}
\maketitle

Taking the square-root of the Klein-Gordon Hamiltonian has opened up an entirely new realm of physics~\cite{Peskin1995}: the resulting Dirac operator provided a relativistic quantum description of massive spin-$1/2$ fermionic particles, thus disclosing the fine-structure spectra of atoms or the anomalous Zeeman effect~\cite{Peskin1995}. Interestingly, as it emerged much later, Dirac's Hamiltonian has found numerous realizations in a completely different context: in condensed-matter physics it describes various complex materials, such as Graphene~\cite{Wallace1947,CastroNeto2009}, transition-metal dichalcogenide monolayers~\cite{Mak2010,wang2012electronics}, 3D Weyl semimetals~\cite{Armitage2018}, the Bogoliubov-de-Gennes quasiparticles dispersion in superconductors~\cite{de2018superconductivity}, and, more recently, topological insulators (TIs)~\cite{hasan2010colloquium,qi2011topological,ozawa2018topological}.

TIs – a new phase of matter – have to date seen a variety of manifestations, with prominent examples such as the 2D~\cite{Klitzing1980} and 4D quantum Hall effects~\cite{zhang2001four}, 1D topological superconductors~\cite{Kitaev2001}, 2D~\cite{Kane2005}and 3D TIs~\cite{Hsieh2008}, crystalline and quasi-crystalline TIs~\cite{fu2011topological,kraus2012topologicalpump}, and higher-order TIs~\cite{benalcazar2017electric}. All available realizations of TIs, however, share a common feature: their spectral bands are attributed with a nonlocal topological index that is quantized~\cite{hasan2010colloquium,qi2011topological,ozawa2018topological}. Hence, whereas different realizations of TIs can vary locally, as long as their topological characterization persists, they will exhibit the same topological phenomena. In other words, the quantization of topological indices lies at the foundation of the characteristic robustness of bulk responses and associated boundary-phenomena in TIs. Currently, there is no known setting that allows for TIs with bands possessing non-quantized topological indices to present such robust phenomena.

In our work, we theoretically devise and experimentally demonstrate a new type of topological insulator that exhibits non-quantized bulk topology alongside robust boundary states. Within our paradigm, the bands of the system do not admit quantized topological invariants, but the system holds a spectral symmetry such that by taking the square of its Hamiltonian quantized topological indices emerge. Notably, whereas taking the square of the Dirac operator leads to the topologically trivial Klein-Gordon model, in our case, the degenerate bands of the squared model exhibit non-Abelian topology. These generate the topological characterization of the square-root model, as well as its corresponding boundary phenomena.

\begin{figure}[t!]
	\includegraphics[width=\columnwidth]{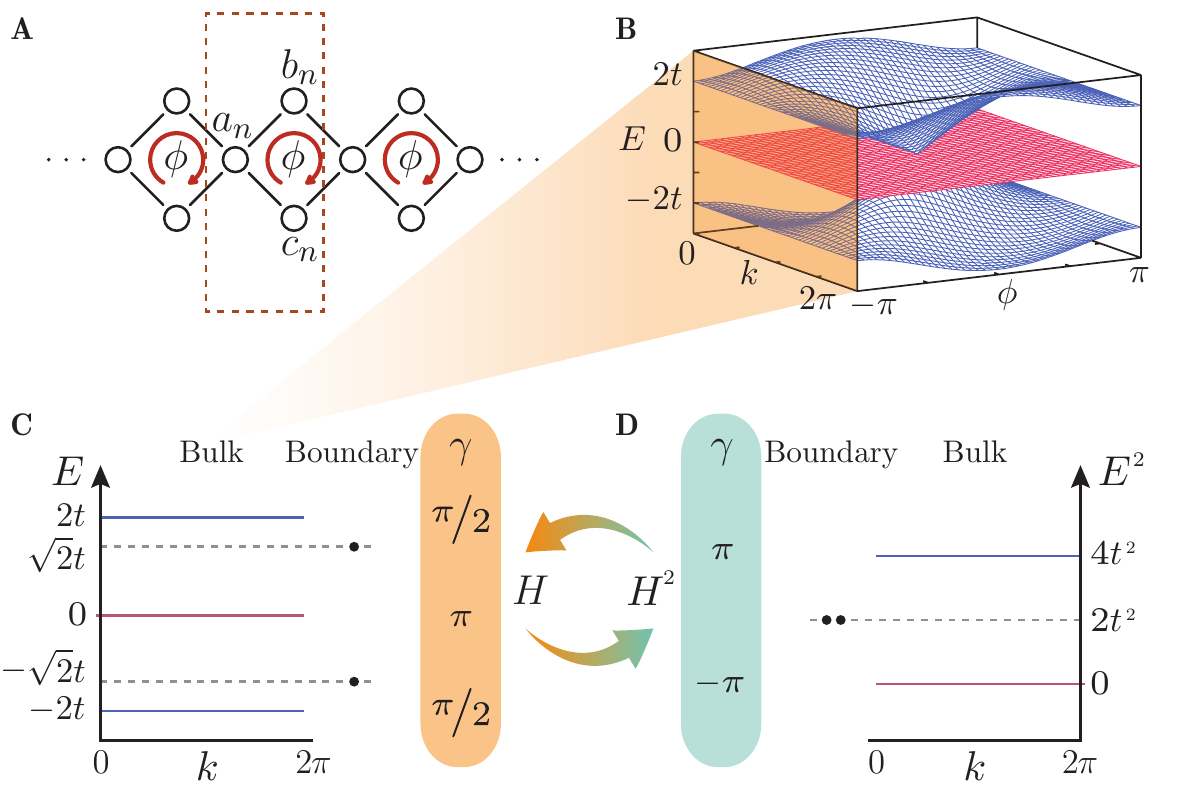}
	\caption{\textbf{Illustration of a square-root topological insulator.} A. The Aharonov-Bohm cages chain [cf. Eq.~\eqref{eq:Hamiltonian mom}], with three elements $a_n, b_n, c_n$ in the $n$th unit cell and a flux $\phi$ threading each plaquette. B. The energy dispersion $E(k)$ of the chain as a function of the flux $\phi$. C. The energy dispersion $E(k)$ at $\phi=\pi$ consists of three flat band at energies $0, \pm 2t$. The band at $E=0$ exhibits a quantized Zak's phase $\gamma=\pi$ while the other two bands show a non-quantized winding of $\pi/2$. At a termination of the chain with an inter-plaquette mode $a_n$, two in-gap boundary states appear at $E=\pm \sqrt{2}t$. D. Squaring the Hamiltonian \eqref{eq:Hamiltonian mom} yields a model \eqref{eq:Hamiltonian squared} with one flat band at $E=0$ and two degenerate flat bands at $E=4t^2$. Both bands exhibit a quantized non-Abelian Wilzcek-Zee phase $\gamma=\pm \pi$. \label{fig1} }	
\end{figure}

We demonstrate our general paradigm using a quasi-1D chain made of photonic Aharonov-Bohm cages. Optical settings prove to be ideally suited for realizing various topological phenomena~\cite{ozawa2018topological}, such as Floquet TIs~\cite{rechtsman2013photonic}, TIs on a silicon platform~\cite{hafezi2013}, 2D~\cite{kraus2012topologicalpump} and 4D topological Hall physics~\cite{zilberberg2018photonic}, as well as non-hermitian topological physics \cite{weimann2017topologically}. In this vein, we employ photonic waveguide lattices with effective negative hopping amplitudes for realizing a square-root topological insulator.

We consider a chain made of Aharonov-Bohm cages, i.e., a quasi-1D lattice composed of inter-connected plaquettes, see Fig.~\ref{fig1}A. Each lattice site is coupled to its neighbours with hopping amplitude $t$, while each plaquette is threaded by a flux $\phi$. The momentum space Hamiltonian of this model is given by
\begin{align}
H(k) = \left(\begin{array}{ccc}
0 & t + t e^{i k}& te^{-i \phi} + t e^{i k}\\\\
t + te^{-i k}& 0& 0\\\\
te^{i \phi} + t e^{-i k}&0& 0
\end{array}\right).
\label{eq:Hamiltonian mom}
\end{align}
The spectrum of $H(k)$ contains three bands: a central band that remains non-dispersive for all values of the flux $\phi$, and two additional bands that are spectrally symmetric around $E=0$. For $\phi=0$, the three bands cross, while for $\phi = \pi$ the spectrum is gapped with three flat bands at energies $\lambda_i \in \{-2 t,0,  2 t\}$, see Fig.~\ref{fig1}B. The latter case corresponds to the Aharonov-Bohm caging effect, where the particles become immobile due to destructive interference~\cite{vidal1998aharonov,longhi2014aharonov}.

\begin{figure}[t!]
	\includegraphics[width=0.7\columnwidth]{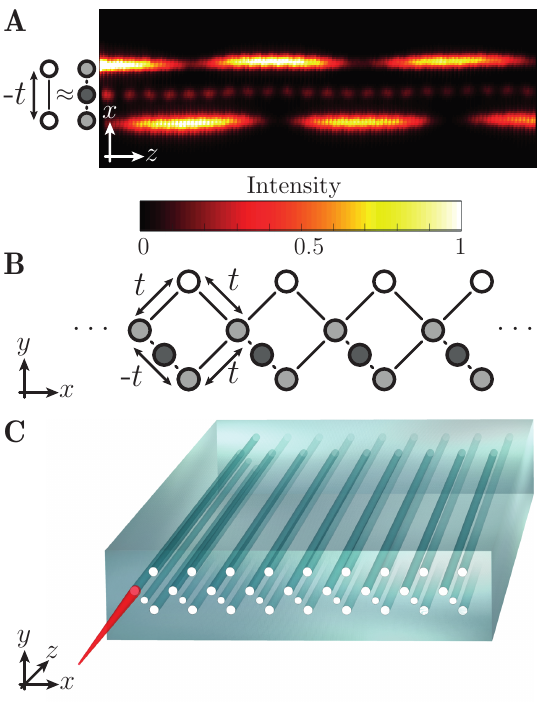}
	\caption{\textbf{Experimental implementation of a non-vanishing flux in each plaquette.} A. By placing an auxiliary waveguide with carefully chosen refractive index profile in between two waveguides, their effective inter-site hopping obtains a phase-shift of $\pi$, i.e., $t\rightarrow -t$. B. Placing such a defect within each plaquette of the lattice structure generates a total flux of $\phi=\pi$.\label{fig2}. C. An illustration of the quasi-1D array of evanescently-coupled waveguides used in the experiment. Light is selectively injected into an input facet of the device and directly imaged using fluorescence microscopy. }	
\end{figure}

For each band, we can evaluate a 1D topological invariant -- Zak's winding phase, $\gamma_i = \int\limits_{\text{BZ}} dk \mathcal{A}_i (k) $, where $\mathcal{A}_i (k)= i\bra{v_i (k)}\partial_k \ket{v_i (k)}$ is the Berry connection of the $i^{\rm th}$ band, and $\ket{v_i (k)}$ is the corresponding eigenstate~\cite{zak1989berry}. The winding phase commonly takes on a binary quantized value of $\pi$ (or $0$) corresponding to encircling (or not-encircling) a singularity in quasi-momentum phase space~\cite{Asboth2015}. Interestingly, we find that the zero-energy band exhibits a quantized winding $\gamma_2 = \pi$, whereas the top and bottom bands display a non-quantized phase $\gamma_1 = \gamma_3 = -\pi/2$. The latter are indeed not quantized, as can be seen by their response to on-site potentials that continuously scramble $\gamma_1$ and $\gamma_3$ and can even make each of them vanish (see supplementary text for more details).

At the same time, the AB-cages chain with open boundary conditions exhibits two localized boundary states at energies $\pm\sqrt{2}t$ on each side of the system, see Fig.~\ref{fig1}C. These boundary states are robust against on-site potentials that do not close the bulk bands, see supplementary text. Commonly, robust topological boundary states appear in a gap that lies above bands that add up to a quantized topological index~\cite{hasan2010colloquium,qi2011topological,ozawa2018topological}. The robust appearance of such boundary states is a surprising occurrence for our case where non-quantized bulk windings arise.

\begin{figure}[t!]
	\includegraphics[width=0.7\columnwidth]{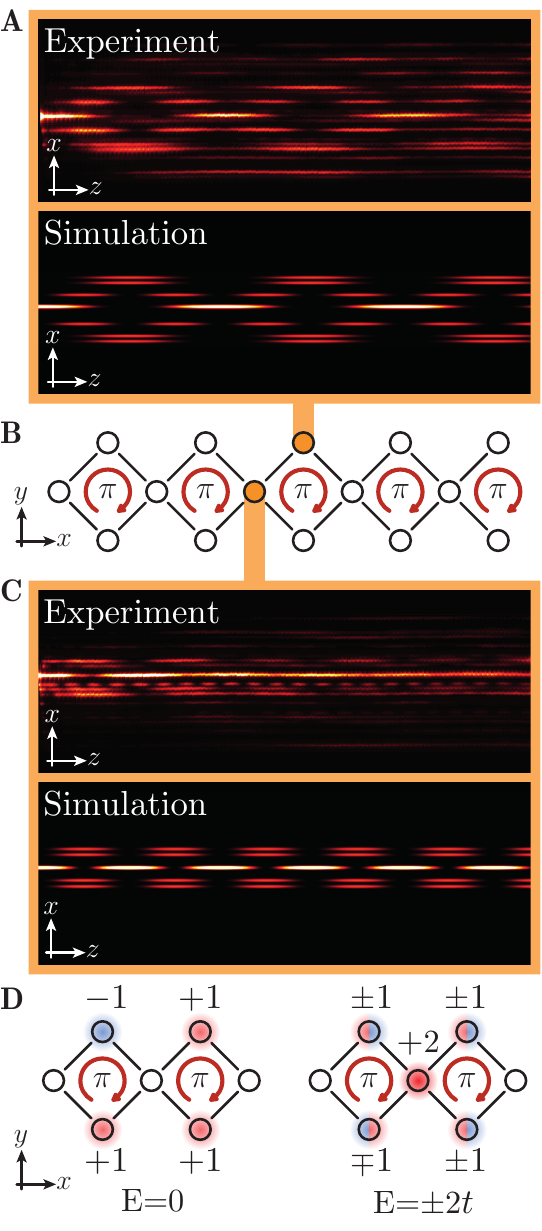}
	\caption{\textbf{Bulk dynamics within the structure.} A. Light dynamics when exciting the top site in a single bulk plaquette. The total envelope remains localized and shows breathing only within the unit cell. B. The two waveguides that are probed in the experiments, in order to demonstrate the flatness of the bulk spectrum. C. Light dynamics in the structure when a waveguide between two plaquettes is excited. The total envelope is also localized and only a local breathing exists. D. The amplitude distribution of the three eigenstates of the system. \label{fig3}}	
\end{figure}

The elusive topological aspects of the model are revealed by taking the square of the Hamiltonian~\ref{eq:Hamiltonian mom},
\begin{align}
H^2 (k) = \left(\begin{array}{ccc}
2m_0 & 0& 0\\\\
0&  m_0 + m_1 (k)& i m_2 (k)\\\\
0&-i m_2 (k)& m_0 - m_1 (k))
\end{array}\right)\,,
\label{eq:Hamiltonian squared}
\end{align}
where $m_0 = 2t^2$, $m_1 (k) = 2t^2 \cos(k)$ and $m_2 (k) = 2t^2 \sin(k)$. The squared Hamiltonian is block diagonal with a single band $\ket{w_1}$ at energy $\Lambda_1=4t^2$ that is decoupled from the other two states. The latter form a topologically non-trivial 1D model that is mappable to the Su-Schrieffer-Heeger (SSH) model by a rotation of the $2\times 2$ block by $e^{i \sigma_y \frac{\pi}{4}}$, i.e., the second block is equivalent to a SSH chain with $0$ intra-cell coupling, $2t^2$ inter-cell coupling, and a constant $2t^2$ energy shift, see supplementary text. The two bands, $\ket{w_2}$ and $\ket{w_3}$, of the effective SSH sub-model are found at energies $\Lambda_2=0$ and $\Lambda_3=4t^2$, respectively.

Importantly, the bands $\ket{w_1}$ and $\ket{w_3}$ form a degenerate subspace at energy $\Lambda_1=\Lambda_3=4t^2$, thus barring our ability to evaluate an Abelian Zak phase for either of these bands. Instead, the bands of the squared Hamiltonian can be assigned a non-Abelian topological phase using the Wilzcek-Zee formulation which generalizes Zak's phase to multiband scenarios~\cite{wilczek1984appearance},
$\gamma = \int\limits_{k}^{k+2\pi}\text{Tr}\Big(\mathcal{A}(k)\Big)\mathrm{d}k$, where $\mathcal{A}(k)^{nm} =\langle v_n(k)|\partial_k|v_m(k)\rangle$, and $n,m$ run over the involved states. For our model, the Wilzcek-Zee phase of both the zero-energy band and the degenerate subspace is $\pi\,\text{ mod }\,2\pi$, corresponding to a nontrivial TI, see supplementary text. Consequently, the standard bulk-boundary correspondence of 1D TIs applies~\cite{resta2007theory}, and the squared Hamiltonian \eqref{eq:Hamiltonian squared} maintains topological boundary modes in the middle of its gap, i.e., at energy $2t^2$. Note that unlike the standard SSH model and in similitude to the Haldane chain~\cite{pollmann2010entanglement}, at proper termination, two degenerate states appear on each boundary of our model, see supplementary text. Crucially, under the square-root operation these boundary states are mapped onto specific boundaries of $H$, giving rise to the topological boundary states found in the spectrum of $H(k)$, see supplementary text.

We implement the AB-cages chain \eqref{eq:Hamiltonian mom} in photonic waveguide lattices fabricated using the femtosecond laser writing technique in bulk glass~\cite{szameit2010discrete}, see methods. The evolution of light propagating along the $z$-direction of an array of single-mode waveguides can be well-described in the paraxial approximation through a set of coupled mode equations $i \partial_z \psi= H \psi$. The wave function $\psi$ represents the excited optical wavepacket as a superposition of bound modes of the waveguides. The matrix $H$ has diagonal elements corresponding to the refractive indices of the waveguides and off-diagonal coupling elements being proportional to the overlap between the bound modes of neighboring waveguides. Thus, discrete Schr{\" o}dinger equations can be simulated in waveguide arrays with the benefit that the time coordinate in the quantum regime is mapped onto a spatial propagation distance in the optical system. In other words, the propagation of an optical wavepacket through a waveguide system simulates the temporal dynamics of an electron in a potential landscape. Notably, using fluorescence microscopy we can directly image the light propagation along the device~\cite{szameit2007quasi}.

In order to generate an effective AB-phase threading each cage, we use Peierls' substitution and associate an effective phase to one of the hopping amplitudes, see Fig.~\ref{fig2}. Engineering a hopping phase for photons is challenging due to the positive refractive index of each waveguide, and consequently the coupling between the waveguides is always real and positive. Nevertheless, by positioning an auxiliary waveguide with a well-tuned refractive index in between two waveguides~\cite{keil2016universal}, an effective negative coupling between the two original waveguides is generated, see Fig.~\ref{fig2}A and the method section. Crucially, the auxiliary waveguide is engineered such that it does not contribute significantly to the dynamics of the system. Therefore, by placing one (or three) negative couplings in each plaquette of our waveguide structure [see Fig.~\ref{fig2}B], an overall flux of $\pi$ within the plaquettes is created, resulting in the desired AB-caging effect, see Fig.~\ref{fig2}C for an illustration of the device.

We start by establishing our ability to generate the AB-caging effect in the bulk of the chain, i.e., we probe the light dynamics in the bulk of the lattice in order to test the flatness of the bands, see Fig.~\ref{fig3}. Exciting a single waveguide within a plaquette will excite all $k$-states of Bloch bands that overlap with this site. At the same time, if the bands are flat the light will stay bound to the injection point and will not disperse. We perform two experiments corresponding to two different injection sites within the unit cell, see Figs.~\ref{fig3}A-C. Indeed, in both experiments, despite of some residual spreading, the propagating wavepacket remains confined to the injected unit cell. The experiments agree well with tight-binding simulations of the AB-chain~\eqref{eq:Hamiltonian mom}.

Interestingly, from the experimental results shown in Fig.~\ref{fig3}C, we can additionally detect the energy of the bands: launching light into a waveguide that connects two plaquettes (Fig.~\ref{fig3}C) solely excites the two modes in the bands at $E=\pm 2t$, as the mode from the band at $E=0$ has no weight in this site, see Fig.~\ref{fig3}D. The resulting beating pattern is, therefore, generated by two modes with a beating length $l_b$ that is connected to the energy difference $\Delta E$ of the participating modes by~\cite{yariv1989quantum} $l_b = \frac{\pi}{\Delta E}$. From the beating in Fig.~\ref{fig3}C, we measure $l_b=0.9$cm, which corresponds to $\Delta E=\pm 3.4$ cm$^{-1}$. Taking into account the spectral symmetry of the model, the energy of the two bands are therefore measured to be at $E=\pm 1.7$ cm$^{-1}$ while the third band lies at $E=0$.

\begin{figure}[t!]
	\includegraphics[width=0.7\columnwidth]{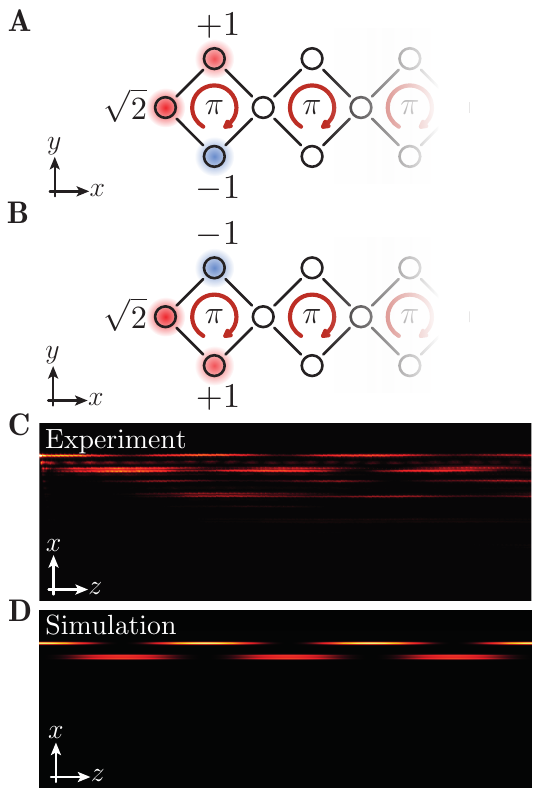}
	\caption{\textbf{Probing the topological edge states.} A. Amplitude distribution of the edge mode at $E=\sqrt{2}t$. B. Amplitude distribution of the edge mode at $E=-\sqrt{2}t$. C. Launching light into the outermost waveguide of the structure, both edge states are evenly excited, such that the experimental light field pattern shows a well visible beating between them. D. This behavior is confirmed by tight-binding simulations. \label{fig4}}	
\end{figure}

We now turn to demonstrate the existence of the topological boundary states in our square-root model. The amplitude distribution of the predicted boundary modes is shown in Figs.~\ref{fig4}A and B. The two boundary states differ by a phase flip and appear at two different eigenenergies, cf.~Fig.~\ref{fig1}C. Hence, similarly to the bulk experiments above, light injected into the outermost waveguide simultaneously excites both topological boundary modes and the resulting light pattern exhibits a beating with a frequency corresponding to the difference between their eigenenergies, see Fig.~\ref{fig4}C. Our experimental data agree well with tight-binding simulations shown in Fig.~\ref{fig4}D. From the beating structure, we can determine the energy of the boundary modes $E_e$: we observe a beating with $l_b=1.3$cm and, hence, deduce that $E_e=\pm 1.2$ cm$^{-1}$. Comparing the observed energies in the bulk and in the boundary, we find that  $\frac{E^2}{E_e ^2}\sim 2$, in agreement with the model predictions \eqref{eq:Hamiltonian mom}. In other words, the observed boundary modes are indeed the topological boundary states that originate from the squared model \eqref{eq:Hamiltonian squared}.

We have predicted and demonstrated the physics of a non-quantized square-root topological insulator, using a photonic platform. Specifically, we show that such systems can exhibit bands with non-quantized topological invariants, but with robust in-gap boundary phenomena. Quantized bulk topology is revealed by mathematically squaring the system's model, where a standard non-Abelian topological characterization is available. We establish this fascinating relationship in the bulk of the system and derive the respective bulk-edge correspondence. Using a chain of AB-cages, we have presented a minimal example of the non-quantized square-root topological insulator phase. We, therefore, expect that our work will stimulate a range of new theoretical
and experimental studies exploring the implications and breadth of such topological phases. In this vein, our experimental results prompt various important questions: Can the square-root topological insulator phase be realized in ultracold atomic setups, where topological quantities can be observed via bulk wavepacket dynamics, rather than by detection of boundary states? What happens to this phase in the presence of interactions, whether in optical, atomic, or condensed matter systems? Is there an analogous square-root topological insulator in the quantum many-body regime and, if yes, will such a model exhibit exotic excitations on the boundary, e.g., parafermions~\cite{fendley2012parafermionic}. Are there other non-linear maps between Hamiltonians that admit such a description~\cite{arkinstall2017topological}? The answers to these questions are now in experimental reach.

During the submission of our manuscript, we came across another recent realization of photonic Aharonov-Bohm cages~\cite{Muk2018}.

\textbf{Acknowledgments} I.P. and O.Z. acknowledge financial support from the Swiss National Foundation. A.S. thanks the Deutsche Research Foundation (grants SZ 276/7-1, SZ 276/9-1, BL 574/13-1, SZ 276/19-1).



\newpage
\cleardoublepage
\setcounter{figure}{0}
\renewcommand{\figurename}{Supplementary Material Figure}

\onecolumngrid
\begin{center}
\textbf{\normalsize Supplemental Material for}\\
\vspace{3mm}
\textbf{\large Non-quantized square-root topological insulators:\\a realization in photonic Aharonov-Bohm cages}
\vspace{4mm}

{ Mark Kremer$^{1,\ast}$, Ioannis Petrides$^{2,\ast}$, Eric Meyer$^{1}$, Matthias Heinrich$^{1}$, \\ Oded Zilberberg$^{2}$, and Alexander Szameit$^{1}$}\\
\vspace{1mm}
\textit{\small $^{1}$Institut f\"ur Physik, Universit\"at Rostock, Albert-Einstein-Strasse, 18059 Rostock, Germany.\\
$^{2}$Institut f\"ur Theoretische Physik, ETH Z\"{u}rich, Wolfgang-Pauli-Stra\ss e 27, 8093 Z\"urich, Switzerland.
}

\vspace{5mm}
\end{center}

\setcounter{equation}{0}
\setcounter{section}{0}
\setcounter{figure}{0}
\setcounter{table}{0}
\setcounter{page}{1}
\makeatletter

\renewcommand{\thefigure}{S\arabic{figure}}

\setcounter{enumi}{1}
\renewcommand{\theequation}{S\Roman{enumi}.\arabic{equation}}
\renewcommand{\thesection}{\Roman{section}}

\section{Materials and Methods}
The waveguides were written inside a high-purity 10cm long fused silica wafer (Corning 7980) using a RegA 9000 seeded by a Mira Ti:Al2O3 femtosecond laser\cite{szameit2007quasi}. Pulses centered at 800nm with duration of 150fs were used at a repetition rate of 100kHz and energy of 450nJ. The pulses were focused 500$\mu$m under the sample surface using an objective with a numerical aperture (NA) of 0.35 while the sample was translated at constant speed of 40mm/min, 200mm/min and 220mm/min, corresponding to the different detunings, by high-precision positioning stages (ALS130, Aerotech Inc.). The mode field diameters of the guided mode were 10.4$\mu$m $\times$ 8.0 $\mu$m at 633nm. Propagation losses were estimated to be 0.2dB/cm. The waveguides are equally spaced by 21.5$\mu$m for the positive and 28$\mu$m for the negative coupling, resulting in an inter-site hopping of $\left |t \right |=0.85$cm$^{-1}$.

For the direct monitoring of the light propagation in our samples, we used a fluorescence microscopy technique~\cite{szameit2007quasi}. A massive formation of nonbridging oxygen hole color centers occurs during the writing process, when fused silica with a high content of hydroxide is used, resulting in a homogeneous distribution of these color centers along the waveguides~\cite{dreisow2009polychromatic}. When light from a Helium-Neon laser at $\lambda=633$nm is launched into the waveguides, the non-bridging oxygen hole color centers are excited and the resulting fluorescence ($\lambda=650$nm) can be directly observed using a CCD camera with an appropriate narrow linewidth filter. As the color centers are formed exclusively inside the waveguides, this technique yields a high signal-to-noise ratio.

\section{Winding phases of $H$}
The spectrum of $H(k)$ [cf.~Eq.~(1) in the main text] with an Aharonov-Bohm (AB) flux of $\phi = \pi$ threading the plaquettes has three flat bands at energies $\pm 2 t$ and $0$, see Fig.~1B in the main text. Since the bands are well isolated, we can define an Abelian Zak's phase to each band
\begin{align}
\gamma_i&=\mathrm{i}\int\limits_{-\pi}^{\pi}\mathrm{d}k \langle v_i(k)|\partial_k|v_i(k)\rangle\,,
\label{eq:Zak's phase}
\end{align}
where $i=1,2,3$ labels the bands in increasing energy. The bulk band solutions  $|v_i(k)\rangle$ are found by diagonalizing the Hamiltonian $H(k)$ and can be written in a compact analytical form
\begin{align}
\ket{v_1} = \frac{1}{2\sqrt{2}} \left(\begin{array}{c}
2e^{i k}\\\\
1+e^{i k}\\\\
1-e^{i k}
\end{array}\right)\,,\hspace{10pt}
&
\ket{v_2} = \frac{1}{2}\left(\begin{array}{c}
0\\\\
-1+ e^{i k}\\\\
-1-e^{i k}
\end{array}\right)\,\hspace{10pt},
&
\ket{v_3} = \frac{1}{2\sqrt{2}} \left(\begin{array}{c}
-2e^{i k}\\\\
1+e^{i k}\\\\
1 -e^{i k}
\end{array}\right)\,.
\label{eq:wavefunctions of H}
\end{align}
Using Eqs.~\eqref{eq:Zak's phase} and \eqref{eq:wavefunctions of H}, we find
\begin{align}
\label{eq:windings of H1}
\gamma_1 &= \gamma_3 = \frac{3\pi}{2}\,\text{mod}\,2\pi\hspace{5pt}\\
\gamma_2 &= \pi\,\text{mod}\,2\pi\,.
\label{eq:windings of H11}
\end{align}

\subsection{Symmetry constraints on the winding phases}
The winding phases of $H$ [cf.~Eqs.~\eqref{eq:windings of H1} and \eqref{eq:windings of H11}] are constrained by the symmetries of the AB-chain model with
\begin{align}
\chi=\begin{pmatrix}
1 & 0 & 0\\
0 & -\mathrm{e}^{\mathrm{i}k} & 0\\
0 & 0 & \mathrm{e}^{\mathrm{i}k}
\end{pmatrix}\,,
\end{align}
and
\begin{align}
\Pi=\begin{pmatrix}
1 & 0 & 0\\
0 & \mathrm{e}^{\mathrm{i}k} & 0\\
0 & 0 & -\mathrm{e}^{\mathrm{i}k}
\end{pmatrix}\,.
\end{align}
Applying the symmetries
\begin{align}
\Pi H(k)\Pi^{-1} &= H(-k)\,,\\
\chi H(k)\chi^{-1} &= -H(-k)\,,
\label{eq:symmetries}
\end{align}
leads to the following relations between the winding phases of the bands:
\begin{align}
\Pi:\hspace{20pt} & \gamma_1\hspace{0.75pt} +\hspace{0.75pt} \gamma_3 \in \{0,\pi\} \hspace{32pt}\text{and}\hspace{32pt}\gamma_2 \in  \{0,\pi\}\,,\\
\chi: \hspace{20pt}& \gamma_1 = \gamma_3\in \{0,\pm \pi/2\} \hspace{12pt}\text{and}\hspace{12pt}\gamma_2 \in  \{0,\pi\}\,.
\end{align}
Breaking the $\chi$ symmetry with an on-site potential allows $\gamma_1$ and $\gamma_3$ to take any phase value, but, interestingly, their sum remains a $\mathds{Z}_2$ invariant.

\section{Boundary states of $H$}\label{subsec:Boundary states of H}
\begin{figure}
	\centering
	\includegraphics[width=0.7\linewidth]{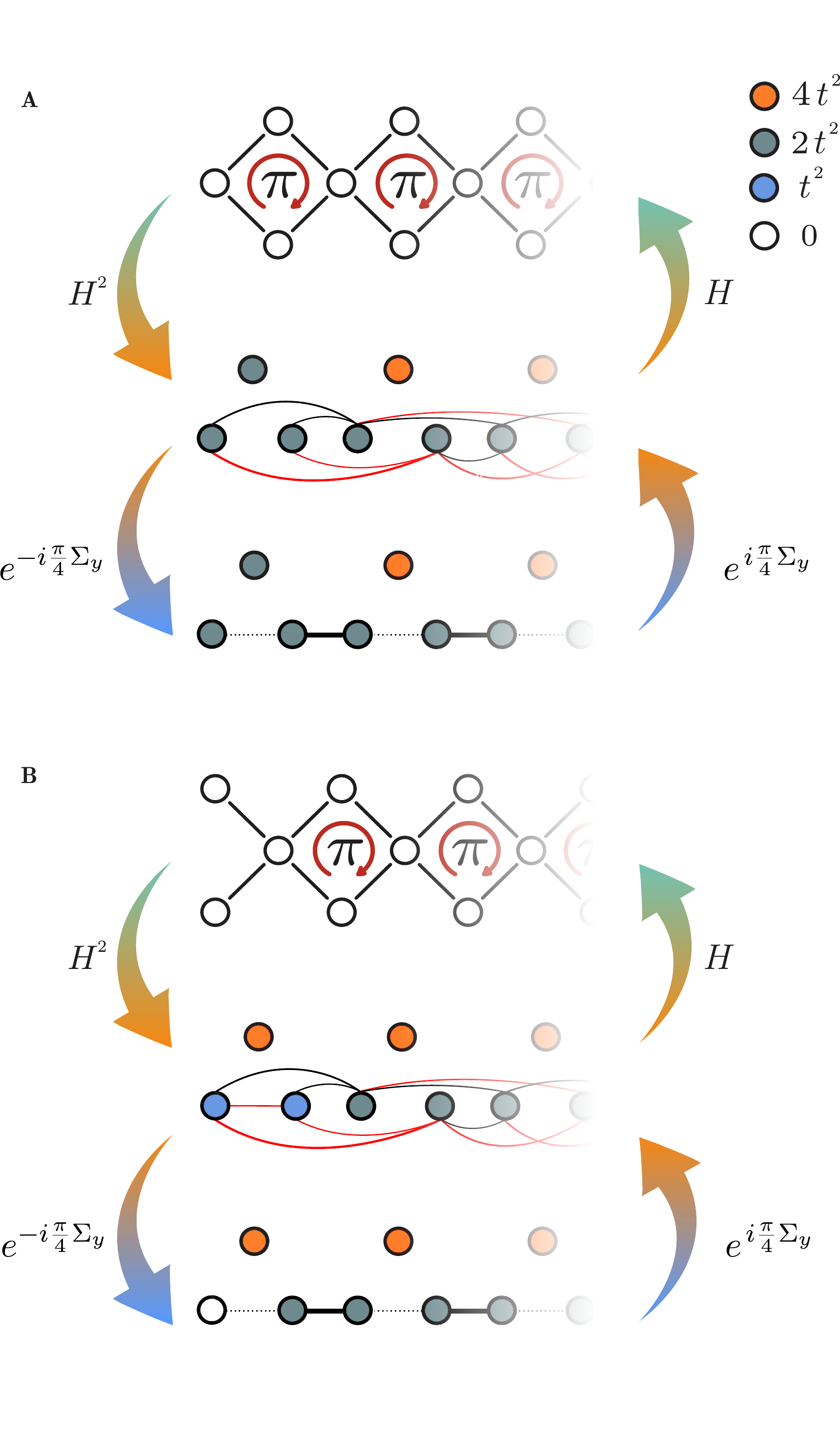}
	\vspace{-20pt}
	\caption{ A. A semi-infinite chain of Aharonov-Bohm cages with a termination that supports two nontrivial topological in-gap boundary modes [cf.~Eq.~\eqref{eq:edge states 1}]. The corresponding $H^2$ model exhibits at its boundary two in-gap degenerate states: one appears as a standard manifestation of the SSH model's bulk-boundary correspondence and the second state is lowered from the bulk's decoupled band of isolated modes [cf.~Eq.~\eqref{eq:Ham2 of edge1}]. B. A semi-infinite chain of Aharonov-Bohm cages with a termination that does not support topological in-gap boundary modes [cf.~Eq.~\eqref{eq:edge states 2}]. The corresponding $H^2$ model similarly does not exhibit topological in-gap boundary states: the standard SSH model has a termination with corresponding boundary mode that is lowered to hybridize with the bulk band at zero energy [cf.~Eq.~\eqref{eq:Ham2 of edge2}]. }
	\label{fig:Edge1}
\end{figure}

\setcounter{enumi}{2}
\setcounter{equation}{0}
To derive the explicit solutions of boundary states, we first consider a semi-infinite AB-cages chain with a termination shown in Fig.~\ref{fig:Edge1}A. The real-space representation of the Hamiltonian is given by
\begin{align}
H=t\begin{pmatrix}
D 		  & 	T	  & 	0	& \cdots\\
T^\dagger &   D		  & 	T	&\ddots\\
0 		  & T^\dagger & 	D	& \ddots\\
\vdots	  &		\ddots&	\ddots	&\ddots
\end{pmatrix}\,,
\label{eq:Ham of edge1}
\end{align}
where $t$ is the hopping amplitude,
\begin{align}
D=\begin{pmatrix}
0 		  & 	1	  & -1\\
1  &   0  & 0\\
-1 & 0 & 0\\
\end{pmatrix}, \hspace{20pt}\text{and} \hspace{20pt} T=\begin{pmatrix}
0 		  & 	0	  & 0\\
1  &   0  & 0\\
1 & 0 & 0\\
\end{pmatrix}\,.
\label{eq:matrices}
\end{align}
Solutions localized within the unit cell of the termination can generally be written as $\ket{e}=\left(a,b,c,0,\cdots\right)^T$ and must satisfy the matrix equation $H\ket{e}=E_e \ket{e}$, with $E_e$ the energy of the boundary state. This leads to two orthogonal solutions given by
\begin{align}
\ket{e_1}=\frac{1}{2}\left(\sqrt{2},1,-1,0,\cdots\right)^T\,\text{and}\,\ket{e_2}=\frac{1}{2}\left(\sqrt{2},-1,1,0,\cdots\right)^T\,,
\label{eq:edge states 1}
\end{align}
with $E_{e_1}=\sqrt{2}t$ and $E_{e_2}=-\sqrt{2}t$.

Following the same procedure, we now consider the alternative termination of the chain, shown in Fig.~\ref{fig:Edge1}B. The real-space representation of the Hamiltonian is given by
\begin{align}
H=\begin{pmatrix}
D 		  & 	T^\dagger	  & 	0	& \cdots\\
T &   D		  & 	T^\dagger	&\ddots\\
0 		  & T & 	D	& \ddots\\
\vdots	  &		\ddots&	\ddots	&\ddots
\end{pmatrix}\,.
\label{eq:Ham of edge2}
\end{align}
Solving the matrix equation $H\ket{e}=E_e \ket{e}$ for a localized state of the form $\ket{e}=\left(a,b,c,0,\cdots\right)^T$ leads to a single solution
%
%
\begin{align}
\ket{e_0}=\frac{1}{\sqrt{2}}\left(0,1,1,0,\cdots\right)^T\,,
\label{eq:edge states 2}
\end{align}
with $E_{e_0}=0$. Importantly, this mode is degenerate with the central bulk band and does not manifest as a topological in-gap state.

\section{Winding phases of $H^2$}
\begin{figure}[h]
	\centering
	\includegraphics[width=0.7\linewidth]{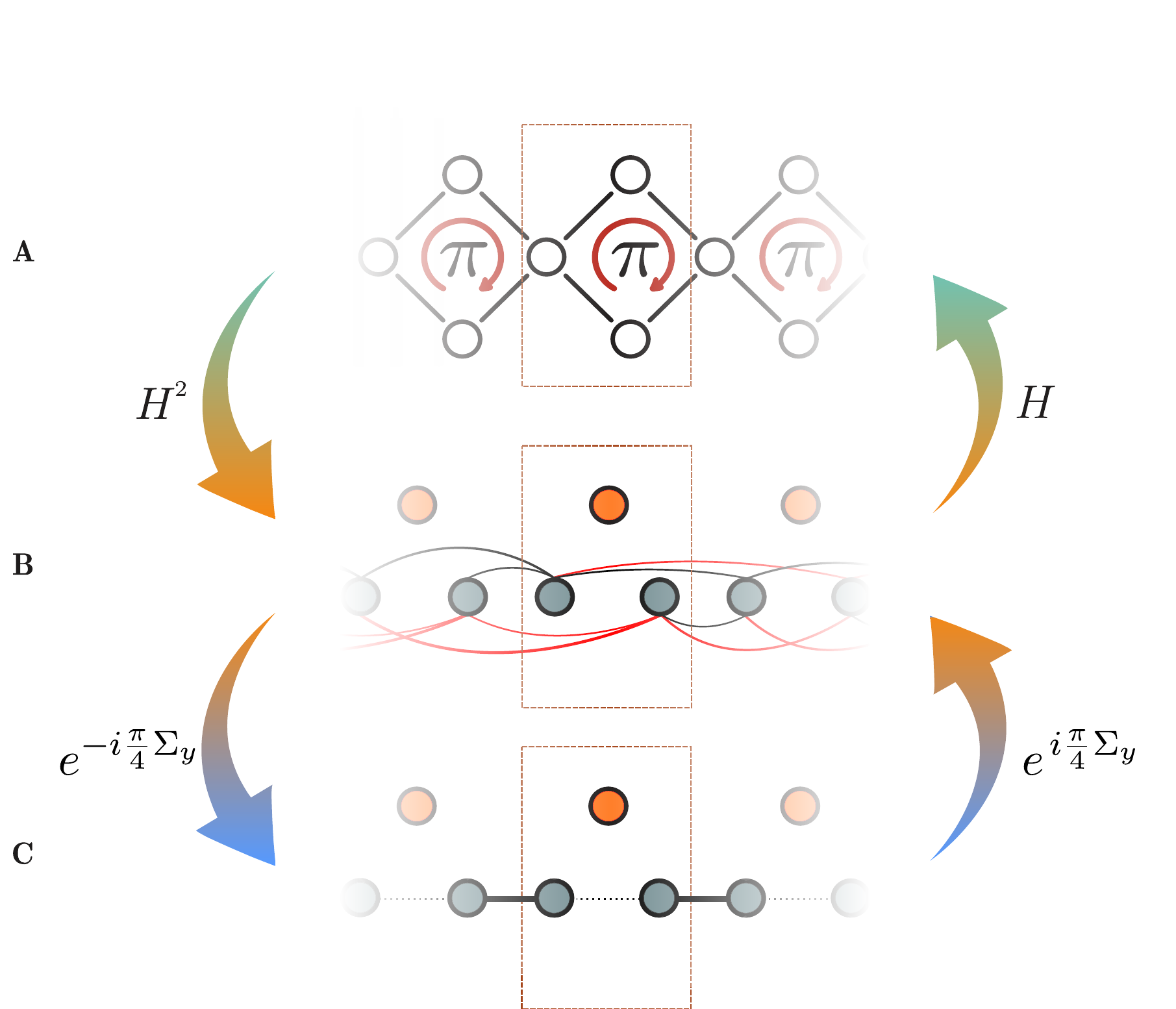}
	\caption{ A. An infinite chain of Aharonov-Bohm cages with $\phi=\pi$ is mapped to B. a chain with positive (black) and negative (red) next-to-nearest neighbour hopping and one decoupled mode per unit cell. Rotating the basis with $e^{-i\frac{\pi}{4}\Sigma_y}$, the latter is mapped to C. the SSH model with an additional decoupled mode per unit cell.}
	\label{fig:bulk-lattice}
\end{figure}

\setcounter{enumi}{3}
\setcounter{equation}{0}
The spectrum of $H^2 (k)$ at $\phi=\pi$, [cf.~Fig.~1D in the main text], has a 2-fold degenerate subspace at energy $4t^2$ and a single band at zero energy. In order to properly capture the non-Abelian nature of the degenerate band we use a generalization of Eq.~\eqref{eq:Zak's phase}, called the Wilzcek-Zee phase,
\begin{align}
\gamma = \int\limits_{k}^{k+2\pi}\text{Tr}\Big(\mathcal{A}(k)\Big)\mathrm{d}k\,,
\label{eq:WZ phase}
\end{align}
where $\mathcal{A}(k)^{nm} =\langle v_n(k)|\partial_k|v_m(k)\rangle$, and $n,m$ run over the involved states. The bulk solutions of $H^2 (k)$ at $\phi=\pi$ can be written in compact form as
\begin{align}
\ket{w_1} = \left(\begin{array}{c}
1\\\\
0\\\\
0
\end{array}\right)\,,\hspace{10pt}
&
\ket{w_2} = \frac{1}{\sqrt{2}}e^{-i \frac{\pi}{4}\Sigma_y }\left(\begin{array}{c}
0\\\\
e^{i k}\\\\
-1
\end{array}\right)\,,\hspace{10pt}
&
\ket{w_3}= \frac{1}{\sqrt{2}}e^{-i \frac{\pi}{4}\Sigma_y }\left(\begin{array}{c}
0\\\\
e^{i k}\\\\
1
\end{array}\right)\,\hspace{10pt} \,,
\label{eq:wavefunctions of H^2}
\end{align}
with eigenvalues $4t^2$, $0$ and $4t^2$, respectively. The $H^2$ model exhibits an inert band composed of decoupled sites, $\ket{w_1}$, that is degenerate with $\ket{w_3}$, see Fig.~\ref{fig:bulk-lattice}A and B for a real-space representation. We have further defined $\Sigma_y =\bigl( \begin{smallmatrix}1 & 0\\ 0 & -\sigma_y\end{smallmatrix}\bigr)$ to be a 3-dimensional matrix related to the $\sigma_y$ Pauli matrix. This rotation maps the $H^2$ model to a generalized Su-Schrieffer-Heeger (SSH) model, see Fig.~\ref{fig:bulk-lattice}C. Using Eq.~\eqref{eq:WZ phase} we find
\begin{align}
\gamma_{1,3}=\pi\,\text{mod}\,2\pi\hspace{5pt}\text{  and  }\hspace{5pt}\gamma_2 = \pi\,\text{mod}\,2\pi\,.
\label{eq:windings of H2}
\end{align}
where $\gamma_{1,3}$ denotes the phase of the degenerate subspace spanned by $\ket{w_1}$ and $\ket{w_3}$, and $\gamma_{2}$ is the phase of the single band at zero energy.

\subsection{Symmetry constraints on the winding phases of $H^2$}
The symmetries obeyed by $H(k)$ are preserved under the squaring operation. The winding phases of $H^2$ are therefore constrained in a similar way using the symmetry transformations
\begin{align}
\Pi H^2(k)\Pi^{-1} &= H^2(-k)\,,\\
\chi H^2(k)\chi^{-1} &= H^2(-k)\,.
\end{align}
This leads to the following relations between the Wilzcek-Zee phases:
\begin{align}
\Pi:\hspace{20pt} & \gamma_{1,3} \in \{0,\pi\} \hspace{12pt}\text{and}\hspace{12pt}\gamma_2 \in  \{0,\pi\}\,,\\
\chi: \hspace{20pt}& \gamma_{1,3}\in \{0, \pi\} \hspace{12pt}\text{and}\hspace{12pt}\gamma_2 \in  \{0,\pi\}\,.
\label{eq:quantisation of H^2}
\end{align}
Importantly, breaking the $\chi$-symmetry with an on-site potential does not make $\gamma_{1,3}$ nor $\gamma_2$ lose its quantization. This is because the $\Pi$-symmetry protects the quantization.

\section{Boundary states of $H^2$ and their mapping to the boundary states of $H$}
\setcounter{enumi}{4}
\setcounter{equation}{0}
The bulk indices of $H^2(k)$ show that the bulk band polarization of the chain is quantized to non-trivial values. However, since the bulk material is insulating, the bulk band polarization manifests as localized states on the boundary~\cite{resta2007theory}. In the following, we give the correspondence between the boundary state found in $H$ (see section~\ref{subsec:Boundary states of H}) and the boundary states found in $H^2$. To this end, we will analyze the two inequivalent lattice termination of $H$ and map those to lattice terminations of $H^2$.

We start by considering the lattice termination of $H$ given in Eq.~\eqref{eq:Ham of edge1} (cf.~Fig.~\ref{fig:Edge1}A). Squaring this matrix leads to
\begin{align}
H^2=\begin{pmatrix}
D^2 + TT^\dagger 		  & 	\{D,T\}	  & 	0	& \cdots\\
\{D,T^\dagger\} &   D^2 + \{T,T^\dagger\}		  & 	\{D,T\}	&\ddots\\
0 		  & \{D,T^\dagger\} & 	 D^2 + \{T,T^\dagger\}	& \ddots\\
\vdots	  &		\ddots&	\ddots	&\ddots
\end{pmatrix}\,,
\label{eq:Ham2 of edge1}
\end{align}
where we have used the fact that $TT=T^\dagger T^\dagger = 0$. The resulting edge termination of $H^2$ is shown in Fig.~\ref{fig:Edge1}A. Importantly, at the edge, the decoupled mode from the isolated bulk band appears at a lower energy. This is due to the fact that, at the edge, the matrix $D^2 + TT^\dagger$ describes the intra-cell hopping, as opposed to the matrix $D^2 + \{T,T^\dagger\}$ in the bulk. Localized solutions of the form $\ket{e^2}=\left(a,b,c,0,\cdots\right)^T$ are found by solving the matrix equation $H^2\ket{e^2}=E_{e^2} \ket{e^2}$. This leads to two orthogonal solutions,
\begin{align}
\ket{e^2_1}=\left(1,0,0,0,\cdots\right)^T\,\text{and}\,\ket{e^2 _2}=\frac{1}{\sqrt{2}}\left(0,1,-1,0,\cdots\right)^T\,,
\end{align}
with $E_{e^2_1}=E_{e^2_2}=2t^2$. We note that $\ket{e^2_2}$ is related to the standard bulk-edge correspondence of the SSH model, while $\ket{e^2_1}$ appears to be originating from a non-topological inert band. Nevertheless, this distinction is basis-dependent and the appearance of the doubly-degenerate boundary of our generalized SSH is a crucial manifestation of the non-Abelian quantized polarization of our model. Under the square-root operation, the boundary states of $H$, given in Eq.~\eqref{eq:edge states 1}, are mapped to linear combinations of $\ket{e^2 _1}$ and $\ket{e^2 _2}$:
\begin{align}
\ket{e _1} = \frac{1}{\sqrt{2}}\left(\ket{e^2 _1} + \ket{e^2 _2}\right)\,\text{and}\,
\ket{e _2} = \frac{1}{\sqrt{2}}\left(\ket{e^2 _1} - \ket{e^2 _2}\right)\,.
\end{align}
%

Following the same procedure as above, we now consider the lattice termination of $H$ given in Eq.~\eqref{eq:Ham of edge2} (cf.~Fig.~\ref{fig:Edge1}B). This leads to the squared Hamiltonian
\begin{align}
H^2=\begin{pmatrix}
D^2 + T^\dagger T 		  & 	\{D,T^\dagger\}	  & 	0	& \cdots\\
\{D,T\} &   D^2 + \{T,T^\dagger\}		  & 	\{D,T^\dagger\}	&\ddots\\
0 		  & \{D,T\} & 	 D^2 + \{T,T^\dagger\}	& \ddots\\
\vdots	  &		\ddots&	\ddots	&\ddots
\end{pmatrix}\,.
\label{eq:Ham2 of edge2}
\end{align}
The resulting edge termination of $H^2$, shown in Fig.~\ref{fig:Edge1}B, displays an intra-cell coupling of the form $D^2 + T^\dagger T $. This leads to a localized solution,
%
%
\begin{align}
\ket{e^2 _0} = \frac{1}{\sqrt{2}}\left(0,1,1,0,\cdots\right)^T\,,
\end{align}
with energy $E_{e^2_0} = 0$. Comparing with Eq.~\eqref{eq:edge states 2}, we find that the above state is mapped onto the same state of $H$:
\begin{align}
\ket{e _0} = \ket{e^2 _0}\,.
\end{align}

In summary, we provide here a mapping between the boundary phenomena of $H$ and $H^2$. The quantized bulk topological winding of $H^2$ leads to a standard topological bulk-edge correspondence reminiscent of the SSH model. The direct relationship between the two models establishes the appearance of topological states in $H$, despite the fact that the square-root model does not have quantized topological indices.

\section{On-site potentials}
\begin{figure}
	\centering
	\includegraphics[width=0.5\columnwidth]{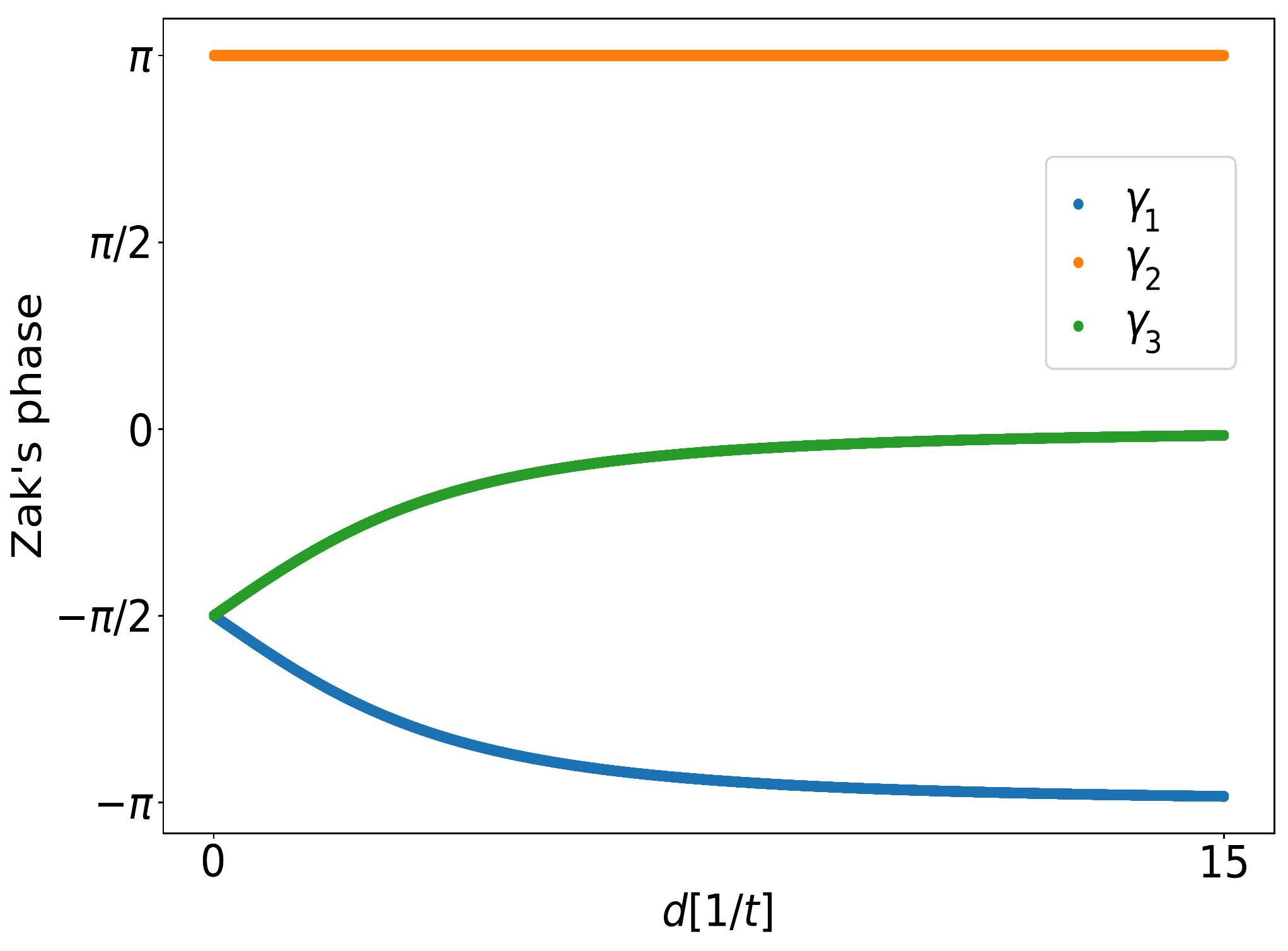}
	\caption{ Zak's winding phases of the bands of $H$ [cf.~Eq.~\eqref{eq:Hamiltonian mom+potential}] as a function of the on-site potential $d$. The winding of the zero-energy band $\gamma_2$ remains unaltered, whereas $\gamma_1$ and $\gamma_3$ are continuously altered to the point that $\gamma_1 \rightarrow \pi$ and $\gamma_3 \rightarrow 0$.}
	\label{fig:windings}
\end{figure}

\setcounter{enumi}{5}
\setcounter{equation}{0}
Adding an on-site potential of the form
\begin{align}
H(k) = \left(\begin{array}{ccc}
d & t + t e^{i k}& te^{-i \phi} + t e^{i k}\\\\
t + te^{-i k}& 0& 0\\\\
te^{i \phi} + t e^{-i k}&0& 0
\end{array}\right),
\label{eq:Hamiltonian mom+potential}
\end{align}
leads to scrambling of winding phases $\gamma_1$ and $\gamma_3$, while $\gamma_2$ remains quantized, see Fig.~\ref{fig:windings}. Importantly, the boundary states given in Eq.~\eqref{eq:edge states 1} remain solutions of the Hamiltonian but with modified energies
\begin{align}
E_{e_1} = \frac{1}{2}\left(d+\sqrt{d^2 + 8}\right)\,\text{ and }\,E_{e_2} = \frac{1}{2}\left(d-\sqrt{d^2 + 8}\right)\,,
\end{align}
while the boundary state of Eq.~\eqref{eq:edge states 2} remains pinned to zero energy, i.e., it does not manifest as a topological in-gap state.

The robustness of the boundary states stems from the fact that the $\Pi$-symmetry, [cf.~Eq.~\eqref{eq:symmetries}], remains a symmetry of both $H$ and $H^2$, while the $\chi$-symmetry is broken. Since the quantization of the Wilzcek-Zee phases [cf.~Eq.~\eqref{eq:quantisation of H^2}] persists under this perturbation, the boundary states are unaffected by such a potential.

\section{Noise analysis}
\begin{figure}[h!]
	\centering
	\includegraphics[scale=1]{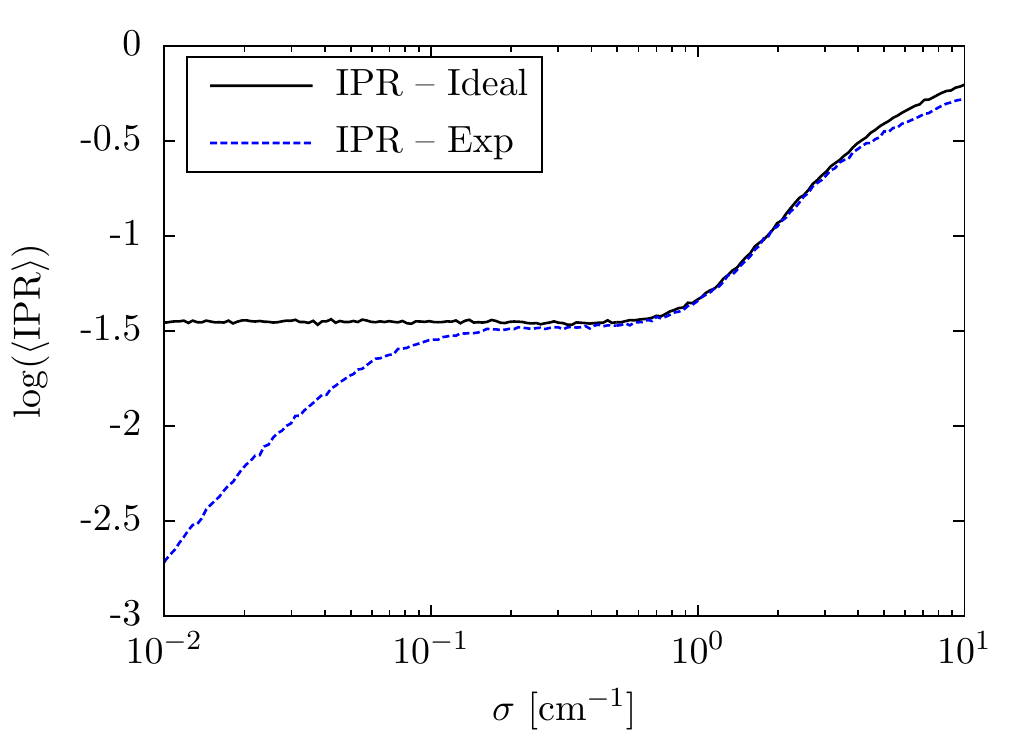}
	\caption{\label{fig_IPRS}The inverse participation ratio (IPR) against the disorder strength $\sigma$. For the simulation we have used 120 lattice sites and every disorder strength was averaged over 100 realizations.}
\end{figure}

\setcounter{enumi}{6}
\setcounter{equation}{0}
In order to determine the effect of noise on the localization of the bulk states, we calculate the inverse participation ratio (IPR), defined as
\begin{equation}
\mathrm{IPR(v)}=\dfrac{\sum\limits_{j=1}^N |v_j|^4}{\left(\sum\limits_{j=1}^N |v_j|^2\right)^2}\,,
\end{equation}
where $v_j$ is the $j$'th component of a state $\ket{v}$. The noise is assumed to be an on-site detuning with zero mean and standard deviation $\sigma$. Since at $\phi=\pi$ all bands are flat, every state is localized within a unit cell, therefore, the IPR is averaged over all eigenstates.
In Fig.~\ref{fig_IPRS} we show the result of this calculation for both the experimental system and our model~\cite{vidal2001disorder}[cf.~Eq.~(1) in the main text]. In both cases the hopping was assumed to be $t=0.77\mathrm{cm}^{-1}$ and every disorder strength was averaged over 100 realisations. Notably, the effect of noise makes the negative coupling a suitable approximation to describe the experimental system.

\end{document}